\makeatletter
\def\input@path{{template/}}
\makeatother
\documentclass[epj,twocolumn]{webofc}
\usepackage[varg]{txfonts}
\woctitle{Emergent Scientist}
\usepackage{amsmath} 
\usepackage{amsfonts}
\usepackage{graphicx} 
\usepackage{setspace}
\usepackage{enumerate}
\usepackage{amssymb} 
\usepackage{hyperref}
\usepackage{float}
\usepackage{wrapfig}
\usepackage{subcaption}
\usepackage{caption}
\usepackage{mathtools}
\usepackage{physics}
\usepackage{pgf, tikz}
\usepackage{tabularx}
\usepackage{xcolor} 
\usepackage{comment}
\usepackage{mhchem}

\begin{document}
\pagenumbering{arabic}
\title{Kinetics of Salt Creeping on a Free Surface: From Nucleation to Saturation }

\author{\firstname{Baptiste} \lastname{Guilleminot}\inst{1}\fnsep\thanks{\email{baptiste.guilleminot@gmail.com}} \and
        \firstname{Élodie} \lastname{Harlé}\inst{2}\fnsep\thanks{\email{elodie.harle67@gmail.com}} \and
        \firstname{Timothée} \lastname{Herbeau}\inst{1}\fnsep\thanks{\email{herbeautim@yahoo.com}}
}

\institute{Université Paris-Saclay, FRANCE
\and
           Université de Strasbourg, 67000 Strasbourg, FRANCE
}

\date{\today}


\abstract{The phenomenon of salt creeping along a free surface remains only partially understood, particularly with respect to its dynamics. In this work, combining a theoretical model with controlled experiments, we identify three distinct kinetic regimes: an initial exponential growth of the height of the crystallized salt deposit on vertical walls, followed by a linear regime, and a final stage where the height saturates while the crystal deposit thickens logarithmically. This unified description makes it possible to follow the macroscopic kinetics of salt growth on a free surface from its nucleation to saturation. In addition, we complement this macroscopic analysis with numerical simulations that shed light on the evolution of the microscopic crystal structure under varying external conditions (humidity and temperature).
}

\maketitle

\section{Introduction}

Table salt (NaCl) is often the first chemical compound encountered by students in introductory chemistry courses. Although its cubic crystal structure may suggest a simple material, the extensive body of research devoted to its physical and chemical properties shows that salt is involved in a variety of non-trivial effects \cite{Chloride_erosion, CEMENT1}. In particular, its strong corrosive properties, combined with its abundance in marine environments and airborne sea spray, pose significant challenges for the durability of man-made structures. In coastal regions, salt can accumulate on building materials such as metals and concrete, accelerating corrosion and degradation processes. Furthermore, when saline water infiltrates porous materials, subsequent evaporation leads to salt crystallization within the pore space. The growth of these crystals generates mechanical stresses that can progressively damage materials and compromise structural integrity \cite{concrete_attack}.  
Among the various phenomena associated with salt crystallization, this work focuses on a particularly intriguing one: the spontaneous creeping of salt crystals along free surfaces. While threatening buildings' integrity, salt creeping also has practical applications such as solar-driven water desalination: \cite{application-creeping-solar, application-creeping-solar-2}.\\

As brine -a $\mathrm{NaCl}$-saturated solution- evaporates, salt crystals form. While some crystals fall to the bottom of the container, others defy gravity and creep upward along partially immersed vertical surfaces. This phenomenon is called "creeping salt". 
Over time, the crystalline layer deposited onto the vertical surfaces both thickens and extends to greater heights in an irregular manner, ultimately giving rise to cauliflower-like structures, commonly referred to as efflorescent creeping. It was first studied over a century ago by Washburn \cite{washburn2002creeping}.

The vertical crystallization, i.e. the formation of the first thin crystalline layer, starts at the contact line. Indeed, as evaporation proceeds, the contact angle between the solution and the free surface decreases until it reaches a critical value below which crystallization is triggered \cite{qazi2019salt,hazlehurst}. Crystals nucleate within droplets of supersaturated solution, which are more frequent near the contact line because the enhanced local curvature increases the evaporation rate as described by the Kelvin equation \cite{kelvin}. \\

After the initial crystals have formed, the thin crystalline layer spreads vertically along the vertical surface. The exact mechanism driving this upward propagation remains debated \cite{velasco2016evaporation, washburn, hazlehurst}. 
One possible explanation involves capillary rise within the narrow interstices between the salt layer and the surface, in accordance with Jurin’s law \cite{jurin}, which sustains evaporation and promotes crystallization at the upper edge of the film \cite{washburn}. Another explanation is the adhesion of the crystals to the surface. The crystal growth is then maintained by evaporation of the surrounding solution and the wetting of a fresh surface as crystals enlarge \cite{hazlehurst}.\\

Among the parameters that govern the kinetics of this phenomenon, relative humidity (RH) plays a central role. It is defined as the ratio of $p_{\mathrm{v}}$, the water vapor pressure in the air, to $p_{\mathrm{vs}}(T)$, the saturated vapor pressure at temperature $T$: 
\begin{equation*}
{\rm RH} = \frac{p_{\mathrm{v}}}{p_{\mathrm{vs}}(T)} \times 100\%.
\end{equation*}
 
Together with temperature -which partly controls RH, because of the strong dependence of $p_{\mathrm{vs}}$ \cite{rankine_formula}- they control the evaporation and crystallization rates.\\
 
 In this paper, we develop a theoretical model that both explains our experimental observations and captures the full macroscopic kinetics of salt creeping on a free surface at low relative humidity, from the initial nucleation to the eventual saturation of the saline deposit. In parallel with the macroscopic model, numerical simulations were developed to explore possible microscopic mechanisms and their dependence on environmental parameters.
 

\section{Method}
\subsection{Experimental Setup}
\label{experimental_setup}
\begin{figure} 
\centering
    \includegraphics[width=0.7\linewidth]{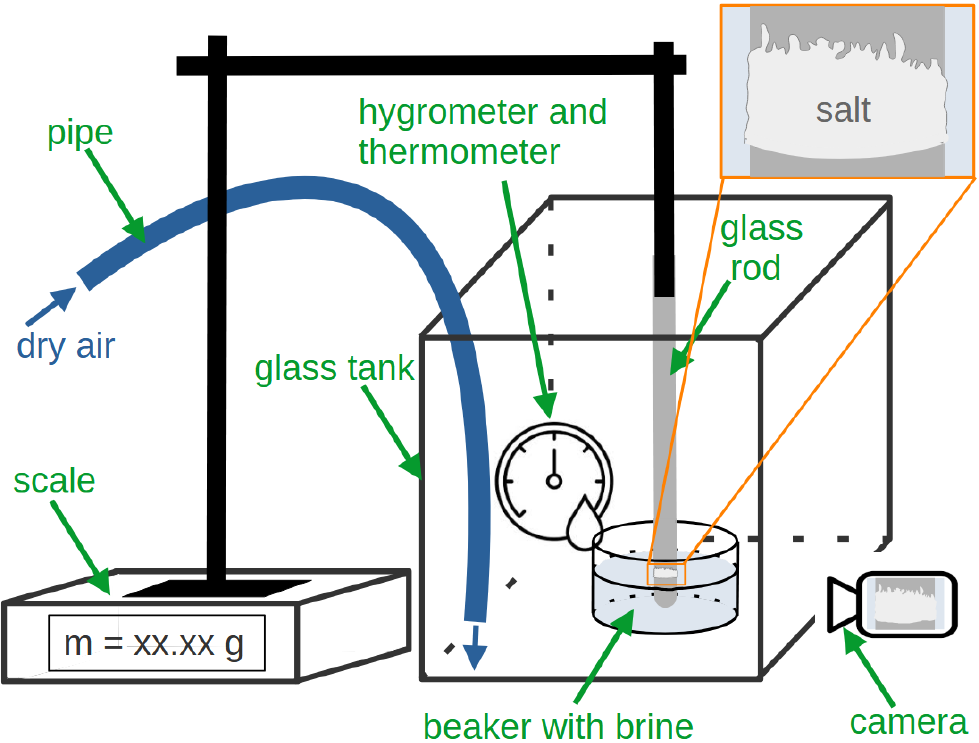}
    \caption{Schematic of the experimental setup. A glass rod is partially immersed in a saturated NaCl solution. To maintain relatively constant experimental conditions (measured by a thermometer and a hygrometer), the system is enclosed in a ventilated box, with dry air injected through a pipe. A camera records the vertical growth of the salt layer. To monitor the mass of salt that has crept up the immersed rod, the rod can be suspended via a mechanical arm placed on a scale.}
    \label{setup}
    \label{RD}
\end{figure}

The experimental setup, shown in Fig.\ref{setup}, consists of a glass tank containing a beaker filled with a salt-saturated solution (distilled water plus table salt at saturation). A glass rod (test tube) previously rinsed with ethanol to remove all greasy residues, was partially immersed in the solution and connected to a scale that recorded its weight every minute throughout the experiment. We controlled the relative humidity, by adjusting the flow of dry air injected through a pipe in the tank. The top of the tank was covered with a plastic sheet perforated at the edges to allow air circulation. RH and temperature were monitored using a hygrometer and a thermometer placed inside the tank. Images were captured with a phone every 90 seconds. We performed four experiments at room temperature, with relative humidites of 25$\%$, 26$\%$, 30$\%$, 32$\%$, over timescales of approximately 50-80 hours (2-3 days). 
Figure ~\ref{height_width_fit20/06plusimg} presents a typical crystal growth sequence obtained during the experiment conducted at 30\% relative humidity. 

To perform microscopic observations, complementary experiments were conducted by replacing the cylindrical rod with a flat microscope slide. These were carried out under slightly different conditions, with a temperature of $(30 \pm 5)^\circ\mathrm{C}$ and a relative humidity of $(20 \pm 5)\%$, in order to obtain well-developed structures more rapidly. Although the kinetics were modified, the same qualitative evolution was observed, namely the formation of an initial thin crystalline layer followed by thickening and three-dimensional structuring. Within experimental uncertainty, changing the substrate geometry did not alter the overall macroscopic behaviour.
In the same experimental conditions, an additional beaker containing saturated brine was left without any immersed rod. Monitoring its mass loss allowed us to estimate the evaporation flux as $J = 0.56 \times 10^{-4} \ \mathrm{kg} \ \mathrm{m}^{-2} \ \mathrm{s}^{-1}$.

\subsection{Data analysis} 
The height and the width of the crystalline salt deposited onto the rod were measured using the pictures taken during the experiment. The outline of the salt deposit (see Fig.~\ref{height_width_fit20/06plusimg}) was manually determined, using a tracking software (ImageJ).

\noindent The height is defined as the maximum vertical distance, measured from the initial solution level, reached by the salt deposit along the rod. The width is defined as the apparent radial extent of the crystalline layer: it is measured as the distance between the rightmost and leftmost visible crystal edges on the rod, from which the rod diameter is subtracted. This value was then divided by two to obtain the average of the larger thickness of the deposit on each side of the rod, assuming uniform thickening around the rod. Both the height and the width were measured using images taken at regular time intervals throughout the experiment. The error bars are constant and correspond to a reading uncertainty, estimated at $\pm$1 mm.

\section{Results}
\label{results}
\subsection{Growth dynamics}
\begin{figure}
\centering
    \includegraphics[width=1\linewidth]{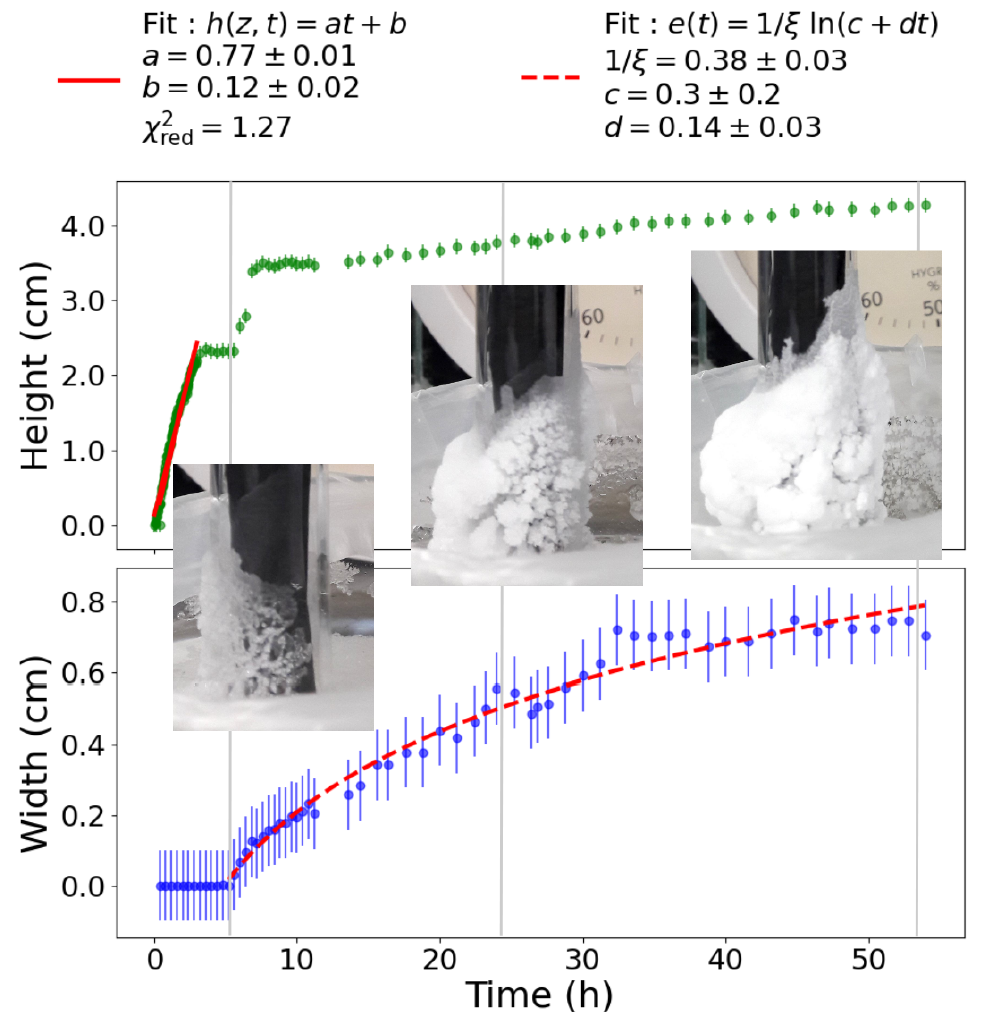}
    \caption{Height (top) and width (bottom) of the crystalline layer creeping on a $16~\mathrm{mm}$ diameter glass rod. The relative humidity was kept at (30 $\pm$ 3) \% and the temperature was (17 $\pm$ 3) °C. The experimental data (symbols) was fitted with the theoretical model (eq. \ref{eq_linear} and \ref{eq_epaisseur} for the height and the width respectively). 
    }
    \label{height_width_fit20/06plusimg}
\end{figure}

Fig.~\ref{height_width_fit20/06plusimg} shows the evolution of the height and of the width of the crystalline film as a function of time. The images taken during the experiment are also shown, illustrating the morphology of the crystalline deposit. Their positions in time are given by the vertical grey lines. \\

As shown in Fig.~\ref{height_width_fit20/06plusimg} (top), the height initially increases linearly during the first five hours, corresponding to the upward spreading of the thin crystalline film. This growth is followed by a plateau, likely caused by a temporary disconnection between the crystalline deposit and the solution. Once the connection is re-established through the formation of new crystals, linear growth resumes —at a rate comparable to that of the initial stage— until approximately $10 \ \mathrm{h}$. Beyond this point, the system enters a new regime characterized by slower vertical growth and enhanced lateral thickening (Fig.~\ref{height_width_fit20/06plusimg} bottom). This transition is visible on the pictures, which show a progressively thicker, whiter film that develops the characteristic three-dimensional cauliflower-like morphology.\newline

In Fig.~\ref{height_width_fit20/06plusimg} (bottom), the temporal evolution of the width is shown. The width begins to increase at the end of the linear growth regime and appears to follow a logarithmic trend, as will be discussed in the following.\newline

For the three other experiments performed at room temperature (with relative humidities of 25\%, 26\%, and 32\%), we observe the same trend: a linear growth of the height (with different slopes) that eventually saturates (this saturation is not observed at the end of the experiment at RH = 25\%), followed by a logarithmic thickening of the salt layer.
\newline

\subsection{Optical microscopy observations}

\begin{figure}
\centering
\includegraphics[width=1\linewidth]{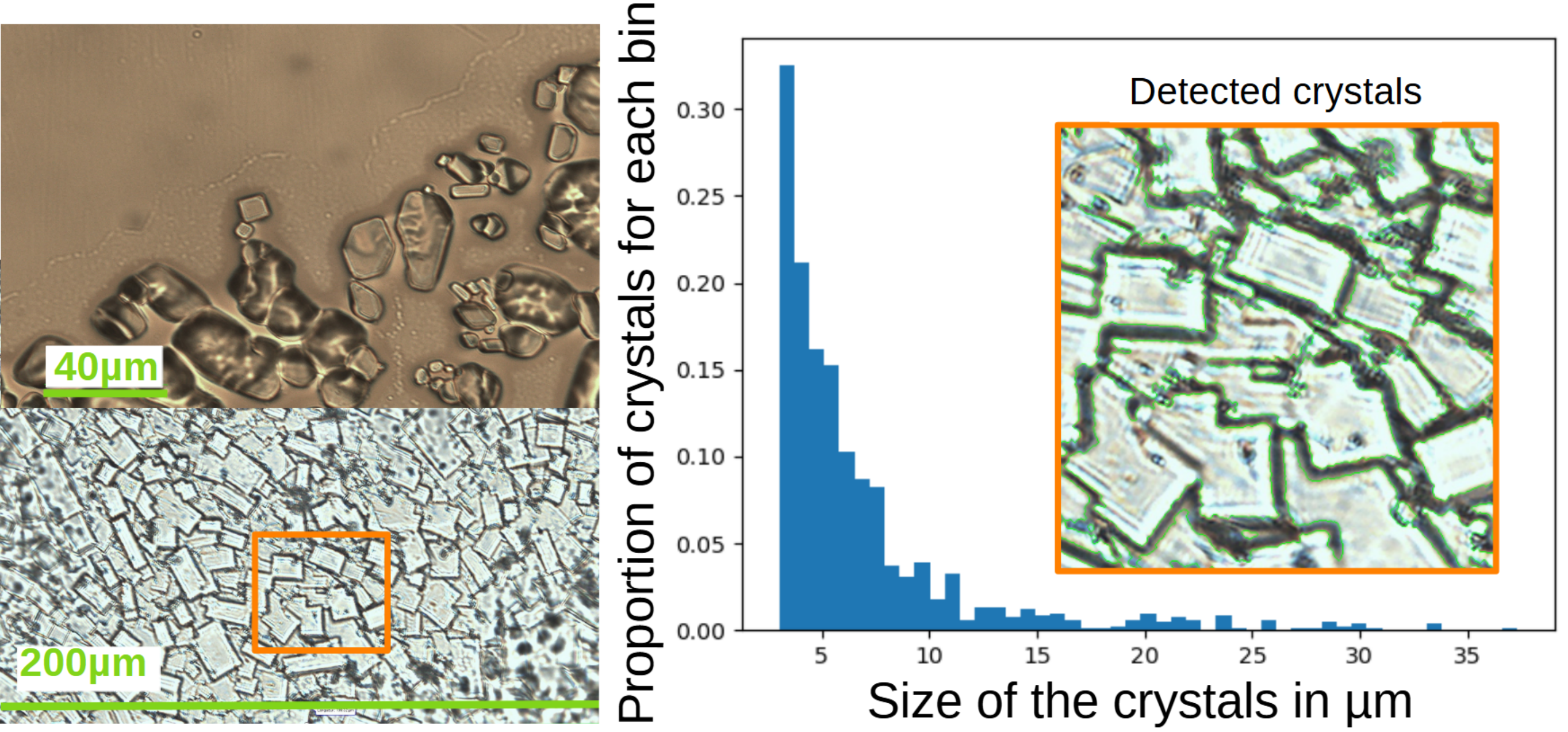}
\caption{Left: Optical microscopy images of the salt layer (image width: $200\ \mathrm{\mu m}$) at the highest point of the crystalline layer (top picture) and approximately $1\ \mathrm{cm}$ lower (bottom picture). Right: Distribution of crystal sizes for the region outlined in orange.}
\label{fig:photo_micro}
\end{figure}

After an experiment conducted around 30 °C and 20\% of relative humidity on a microscope slide (glass), the crystals were observed with a microscope. Examples of the observations are shown in Fig.~\ref{fig:photo_micro} (left). The crystals were then contoured using image analysis software, and their size was estimated as the square root of the area. 
This procedure yields the crystal size distribution shown in Fig.~\ref{fig:photo_micro} (right), which will later be compared with the simulation results (see Section \ref{simulation}).

\section{Theoretical model \& Simulations}
\subsection{Theoretical model}
\label{theoretical_model}

\begin{figure} 
\centering
    \includegraphics[width=0.8\linewidth]{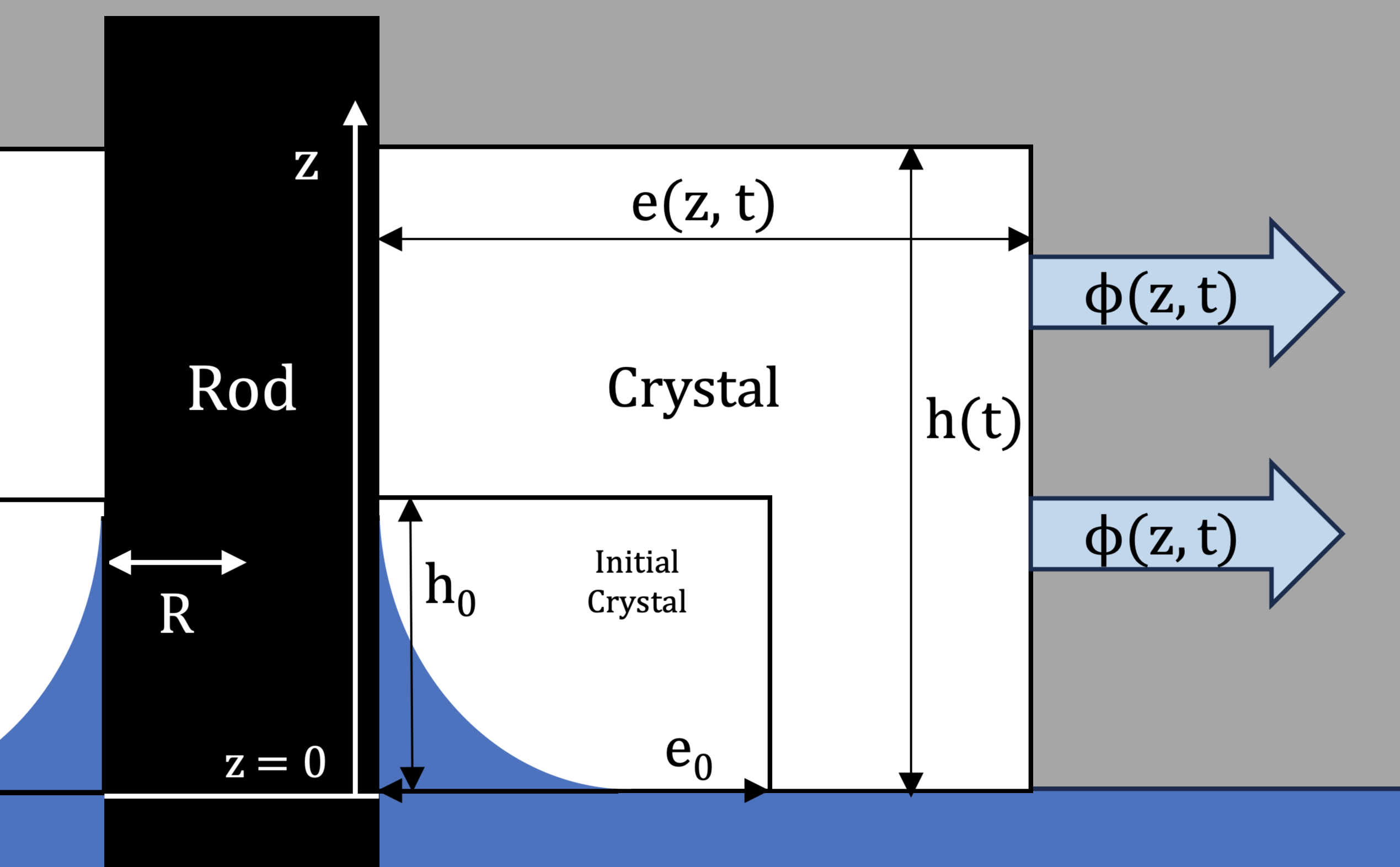}
    \caption{Definition of the different quantities: the evaporation flux $\Phi(z,t)$ of water molecules from the crystalline layer of thickness $e(z,t)$ and height $h(t)$. The first crystal nucleation occurs at the meniscus of brine on the rod of radius $R$. Hence, the meniscus parameters $h_0$ and $e_0$ determine the initial crystal domain. \cite{qazi2019salt}}
    \label{schema_variable}
\end{figure}

We derive an analytical model describing the time evolution of the height $h(t)$ and width $e(z,t)$ of a salt crystalline domain growing on a cylindrical rod, assuming cylindrical symmetry. To that effect, we consider an initial crystal of height $h_0$ and width $e_0$, which subsequently grows in time to a height $h(t)$ and a width $e(z,t)$. These variables are illustrated in Fig.\ref{schema_variable}. Based on experimental observations, Fig.\ref{fig:photo_micro}, we assume that the crystal layer is homogeneously composed of monocrystals of typical volume $\langle V \rangle \approx 10^{-18} \ \mathrm{m}^3$, each containing a number $N_s$ of NaCl molecules. This volume varies little and is taken as constant, except during the nucleation of each crystal.

The ratio $N_e$ of water molecules to NaCl in the solution is fixed by temperature and pressure, as it is at equilibrium. As a consequence, for each water molecule that evaporates at the surface of the crystal layer, $1/N_e$  NaCl molecules crystallize. Assuming that the solution remains saturated in salt, i.e. at $360\  \mathrm{g/L}$, and using usual molar masses we find $N_e=9$. This conservation of flux at the interface, developed in the Supp. Mat, leads to Eq.\ref{Main_equation}:

\begin{align}\label{Main_equation}
       \frac{\rm{d}}{\mathrm{d}t} \int_{0}^{h(t)} e(z,t) \mathrm{d}z = \frac{\langle V \rangle}{N_e N_s} \int_{0}^{h(t)} \Phi(e,z,t) \mathrm{d}z,
\end{align}

Here, $\Phi(e,z,t) $ is the flux of water molecules evaporating per unit surface at the interface between crystal and air (in s$^{-1}$m$^{-2}$).

In the Supp. Mat, in addition to deriving Eq.\ref{Main_equation}, we show that this flux $\Phi(e,z,t)$ can be modelled with three parameters by Eq.\ref{flux} :
\begin{align}\label{flux}
     \Phi(e,z,t) =\Phi_0 \ \mathrm{e}^{-\frac{z}{h_\infty}} \ \mathrm{e}^{-\xi \ e(z,t)},
\end{align}
Where : 
\begin{itemize}
    \item $\Phi_0$: the evaporation flux per unit surface of water molecules at a given pressure and temperature.
    \item $\xi$: the number of $\mathrm{NaCl}$ molecules deposited per unit length, by accretion, as the brine flux passes through the crystal domain. We assume this accretion process to mostly hinder the radial part of the flux, from the central capillary to the external surface of the crystal domain. Indeed, the central capillary is larger than the interstitial space between crystals \cite{washburn} and thus, less impacted.
    \item $h_\infty$ which represents the characteristic height above which gravity hinders the upward transport of brine, preventing it from being supplied to the growing crystal. One could naively think that, if the process was dominated only by gravity, Jurin's law \cite{jurin} would entirely determine the maximal height reached by capillarity: $h_{\mathrm{max}} = 2\gamma \cos(\theta)/ \rho g r\approx 0.7\ \mathrm{m}$, using standard values for water's properties: density $\rho = 10^3 \ \rm{kg.m}^{-3}$, surface tension $\gamma \sim 70\ \rm{mN.m}^{-1}$ and for the wetting angle $\cos(\theta) \sim 1/2$ Fig.\ref{schema_variable}, as well as the typical values for the diameter of the capillaries $r = 10\ \mu \rm{m}$ which we typically observe under the microscope Fig. \ref{fig:photo_micro} . The discrepancy between $h_{\mathrm{max}}$ and $h_\infty$ is due to their definition: the first is a sharp maximum while the second is a characteristic scale in the context of the Boltzmann probability of finding an entity NaCl at a given height at a given temperature. Additional sources of discrepancy with $h_{max}$ come from overlooking the accretion process in the capillaries and the evaporation diminishing the upward flow in the capillaries.
In the following we retain the experimental value $h_\infty \approx 0.1 \ \mathrm{m}$ in order to circumvent the substantial uncertainties associated with the wetting angle, the capillary size, and the value of the surface tension of salt-saturated water. Although the experimental value of $h_\infty$ is used as an input parameter of the theoretical model, it is expected to be independent of the kinetics of salt creeping. This height is determined solely by physical constants and hydrostatic equilibrium.
    
\end{itemize}

We emphasize that the flux in the media (which we equate to the one at interface having a quasi-steady state) cannot be modelled by Darcy's law \cite{Darcy1856}: $ \Phi = \kappa(p_0 - p_{\rm{v}})/(\eta \ e(z,t)) $ with $\kappa$ the permeability, $\eta$ the dynamic viscosity and $p_0 - p_{\rm{v}}$ the difference of partial pressure in water between the outside and the wet crystal domain. The key difference with our model lies in the dependency on $e(z,t)$. 
Darcy’s law is a steady-state description and cannot capture the transient nature of the process. In particular, the crystallization progressively narrows the capillaries, reducing the effective permeability —an effect that Darcy’s law cannot account for. As a result, models based on Darcy’s law predict growth kinetics that are significantly faster than those observed experimentally, which is why such approaches were discarded.

Reynold's transport theorem allows to explicitly differentiate the left hand side of Eq.\ref{Main_equation}:

\begin{align}\label{Global_ODE}
      &\frac{\rm{d}~h(t)}{\mathrm{d}t} \  e(h(t),t)  \\ \nonumber &+ \int_{0}^{h(t)}  \partial_t e(z,t) \ \ \mathrm{d}z =  \frac{\langle V \rangle}{N_e N_s} \int_{0}^{h(t)} \Phi(e,z,t) \ \mathrm{d}z.
\end{align} 
 
We solve Eq.\ref{Global_ODE} under three different sets of approximations, reflecting experimental observations. Three different regimes appear. An initial regime is characterized by free expansion of the crystal. It is followed by an intermediate gravity-driven regime in which vertical growth dominates. Finally, the system enters a saturation regime, where the height plateaus while the crystal thickness $e(z,t)$ in turn increases.\newline

The first regime appears as an exponential evolution of the height. Shortly after the beginning of the experiment  -during the first hour or so- we observed that the thickness remained constant along the rod, while the crystallization height evolved.
As a consequence, we assume the first regime to be defined by:
\begin{align}
    \partial_t e(z,t) \ dz = \dot h \ e_0 \ \delta(z&=h(t)),  & 
    \frac{h(t)}{h_\infty} & \ll 1. 
\end{align}

Under such premises, the height evolution is :
\begin{align}
       h(t) = h_0 \ \exp[ \ {\frac{\alpha e^{-\xi e_0}}{2e_0} t} \  ],
       \label{eq_hauteur}
\end{align}

with $h_0$ the initial height of the crystal formation, and  $\alpha = \Phi_0 \langle V \rangle/(N_e N_s)$. Since the phenomenon is started by the wetting of the rod, $h_0$ is assumed to be the height of the initial brine meniscus: $h_0 = 10^{-3} \ \mathrm{m}$ will be taken to compute characteristic times.\newline 
Similarly, $e_0$ is the thickness of the meniscus. Experimentally, it is equal to the thickness of the first layer of crystallization on the rod -$e_0 = 10^{-3} \ \mathrm{m}$ -, while $1/\xi$ can be interpreted as $e_{\infty}$ the limiting thickness above which the flux cannot go through the media anymore. 
Given our experimental conditions, the numerical values we used are the following: $\Phi_0 = 1.7 \ 10^{22} \ \mathrm{m^{-2}}\mathrm{s^{-1}}$, and $\alpha = 2. 10^{-7} \ \mathrm{m}.\mathrm{s^{-1}}.$ 
Finally, we define a characteristic time $\tau_1  = 2e_0\ \mathrm{e}^{\xi e_0}/\alpha  $ which we will compare to our experimental results.\\

The second regime is the longest one: experimentally, it lasts from several hours to tens of hours, depending on the temperature \& humidity, and is easier to probe. It emerges from Eq.~\ref{Main_equation} under the hypothesis that the thickness is still constant, but the height is no longer negligible compared to $h_\infty$: 

\noindent This gives: $  \partial_t e(z,t) \ dz= \dot h \delta(z=h(t)) e_0, \ \&\ 
    \frac{h(t)}{h_\infty} = \mathrm{O}(1)$.

\noindent This leads to a linear time evolution for $h(t)$ : \footnote{See the details of the calculations in the Supp. Mat} 

\begin{equation}
    h(t) = h_0 + \frac{\alpha}{2e_0}h_{\infty}\ \mathrm{e}^{-\xi e_0} \ t, 
    \label{eq_linear}
\end{equation}. 

From Eq.\ref{eq_linear}, one can extract a characteristic time that is the same as for the first regime. This indicates that a more relevant quantity is $a_2 =h_{\infty}/\tau_2 $, the slope of the linear regime. This $a_2$ is an intrinsic prediction of our model, whereas $\tau_2$ is related to whether the condition $ h(t)/h_\infty =\mathrm{O}(1) $ is valid. The validity of this condition in fact defines the transition between regimes, which is not addressed here. We simply derived a linear time evolution for $h(t)$, in the regime where $h(t)/h_\infty = \mathrm{O}(1) $ and $e(z,t) \sim e_0 $.\\

Finally, the third regime, where $h(t) \sim h_{\infty} $ saturates, leads to a prediction for the width evolution $e[z, t]$. The first term on the left hand side of our main equation \ref{Global_ODE} vanishes and then transforms into: 
\begin{align}
      \alpha e^{-\xi e(z,t)} e^{-\frac{z}{h_\infty}}  - \partial_t e(z,t)   =0.
\end{align} 
\noindent This, in turn, gives the time evolution of the crystal thickness:
\begin{align}
        e(z,t)  = \frac{1}{\xi} \ \ln \left[ C(z) + \xi \alpha e^{-\frac{z}{h_\infty}} \ t \right].
        \label{eq_epaisseur}
\end{align}
 
Similarly to the previous discussion on $\tau_2$ and $a_2$, we do not investigate transitions between regimes, thus, the $z$-dependent though time constant quantity $C(z)$ appearing in the logarithm cannot be derived: there is no clear-cut condition from one regime to another.
However, we insist that our model predicts a logarithmic widening and the relevant quantity we can compare to experiments is a characteristic time $\tau_3 = \mathrm{e}^{\frac{z}{h_\infty}}/(\xi\ \alpha)$ for a given altitude $z$.

\subsection{Simulation} \label{simulation}

\begin{figure}
    \centering
    \includegraphics[width=0.75\linewidth]{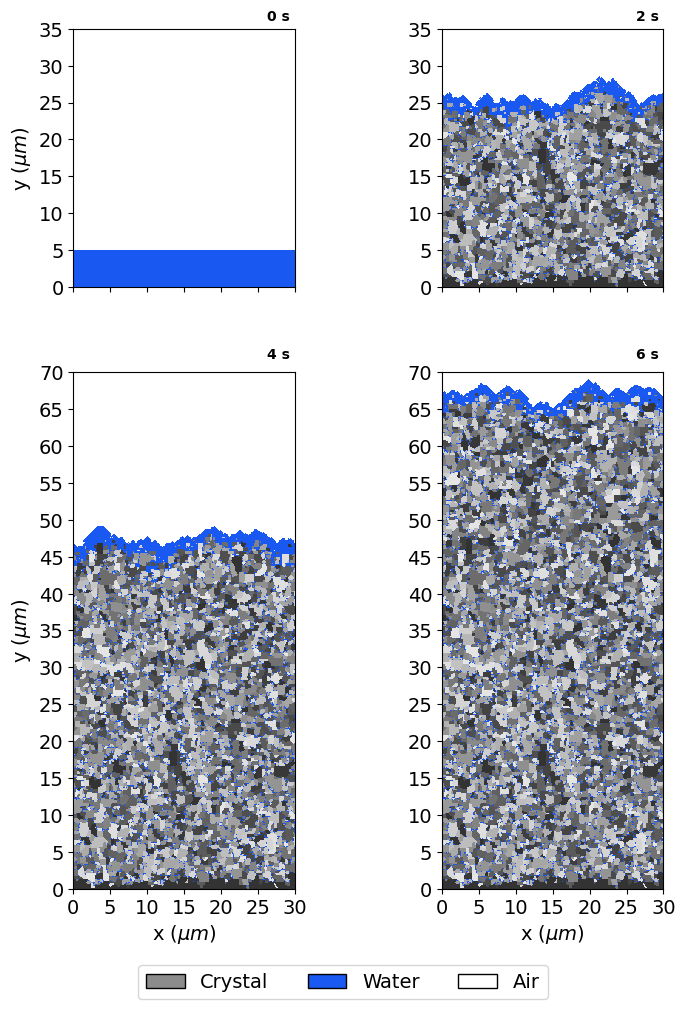}
    \caption{The simulation at different times for a spacing of $\Delta x = 0.1\,\mu\mathrm{m}$, at a temperature of 35 °C. Shades of grey denote different crystals and carry no physical meaning.}
\label{fig:sim_image}
\end{figure}

To simulate mesoscopic crystal growth, we use a two-dimensional cellular automaton defined on a square grid of spacing $\Delta x$ (see Fig. \ref{fig:sim_image}) tangent to the rod. Each cell may be in one of three states: empty, filled with water, or occupied by a NaCl crystal. At $t=0$ a film of brine of thickness $10 \ \mu$m is placed along one edge of the domain; the remainder of the domain is initially dry. Time advances in discrete steps chosen for numerical convenience. Physical time is then recovered from a mass balance using the experimentally measured evaporation rate. At each step, the water configuration is updated through capillary rise, the local salt concentration field is computed from the measured evaporation rate, and crystallization events are sampled from a Boltzmann-type probability distribution incorporating crystal binding energies $E_l=2.63 ~ \mathrm{eV}$, see \cite{kittel2005}. The concentration field is obtained from the Ostwald--Freundlich relation \cite{Ostwald-Freundlich_effect} coupled to a one-dimensional steady diffusion--evaporation model.
Evaporation is prescribed using experimental data: $J = 0.56 \times 10^{-4} \ \mathrm{kg}\cdot \mathrm{m}^{-2}\cdot \mathrm{s}^{-1}$. Further details are provided in the Supplementary Material Sec.~4.

A fully detailed description would require incorporating the wetting angles of water on both salt crystals and glass to determine the capillary rise above the last crystal. For simplicity, we instead impose that the water advances above a crystal of linear size $r$ by a distance proportional to $r$. Additional details are provided in the Supplementary Material.

\begin{figure}
    \centering
    \includegraphics[width=0.7\linewidth]{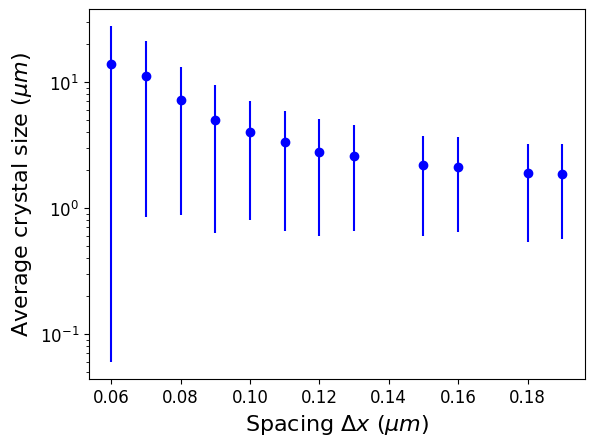}
    \caption{Average size of the crystals with a confidence interval at 68\% as a function of the square grid spacing $\Delta x$.}
    \label{fig:sim_spacing}
\end{figure}

The simulation converges reliably and remains insensitive to the grid spatial resolution (Fig.~\ref{fig:sim_spacing}), except when $\Delta x < 0.1\ \mu$m, where divergence appears because the Ostwald--Freundlich effect must then be included in $\Delta E$.

With a grid spacing of $0.1\ \mu$m, the simulation yields an average crystal size of $3.9\ \mu$m, with the 16th and 84th percentiles at $0.8\ \mu$m and $7.0\ \mu$m, respectively. These values compare with the experimental measurements of $3.1\ \mu$m, $1.1\ \mu$m, and $4.7\ \mu$m (see Fig.~\ref{fig:photo_micro}). While the simulated distribution is somewhat more skewed than observed experimentally, the agreement is overall satisfactory given the approximations inherent in the model.

The simulation yields a growth rate of about $1 \ \mathrm{cm}\cdot\mathrm{h}^{-1}$, in reasonable agreement with the experimental value at the level of order of magnitude, see Tab \ref{tab:time_caract}.

\section{Discussion and Perspectives}
\label{discussion}

By comparing our theoretical predictions (both analytical and numerical) with experimental data, we are able to discuss three regimes presented there. 
Theoretical predictions are based on the values of $h_0 = 10^{-3}$ m, $e_0= 10^{-3} $m, $h_\infty= 10^{-1}$ m, $e_\infty= 5.10^{-3}$ m and $\alpha = \Phi_0 \langle V \rangle/(N_e N_s)$. Each of these depends on phenomena that drive but are not driven by salt creeping -wetting of a surface, evaporation-. We can therefore use them as inputs for the three identified characteristic quantities ($\tau_1$, $\tau_3$ and $a_2$) and compare our predictions to their experimental values, as shown Tab.\ref{tab:time_caract}.

\subsection{First regime - exponential growth}

At the early stage of the experiment (about an hour), we have shown that the height should evolve exponentially as in Eq.~\ref{eq_hauteur}. \\
Probing the predicted height evolution in this first regime is relatively uneasy. The phenomenon starts on a random part of the rod (not necessarily facing the camera) and is spatially inhomogeneous particularly at the beginning. Still, the prediction of our model concerning the characteristic time of this regime was $\tau_1 = 10^{4} \ \mathrm{s}$. The order of magnitude of a few hours is in agreement with the rate of change in our experiments at the given experimental conditions as well as other work \cite{velasco2016evaporation}.

\begin{table}
\centering
\begin{tabularx}{1\linewidth}{|c|X|X|X|}
\hline
 & Regime 1 - $\tau_1$  ($\mathrm{h}$) & Regime 2 -$a_2$ ($\mathrm{cm.h^{-1}}$) & Regime 3  - $\tau_3$ ($\mathrm{h}$) \\ \hline
Theory & $ 4  \pm 1$ & $ 2  \pm1$ & $  10 \pm 2$ \\ \hline
Exp. &  & 0.80 $\pm$ 0.01 &  7 $\pm$ 1 \\ \hline
\end{tabularx}
\caption{Comparison of the characteristic quantities, theoretical and experimental, for the three regimes.}\label{tab:time_caract}
\end{table}

\subsection{Second regime - linear height}

Recall that the second regime starts after the initial transient phase and persists over several tens of hours. In this regime, the height $h$ becomes comparable to its asymptotic value $h_\infty$ (up to $10^{-1}$ m). We obtained a linear increase of the height with time -Eq.\ref{eq_linear}-. Experimentally, we indeed observe this linear growth regime.\\

\begin{figure} 
\centering
    \includegraphics[width=0.7\linewidth]{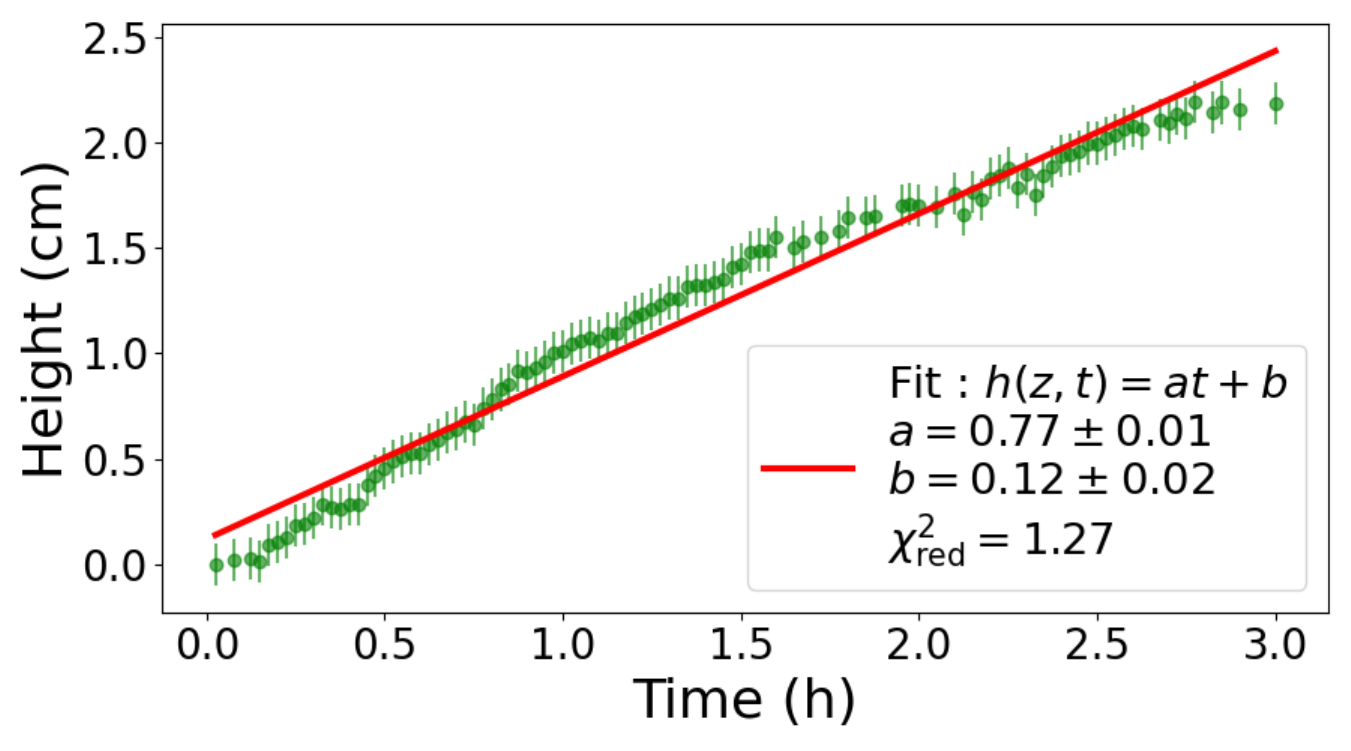}
    \caption{Fit of the experimental height data, focusing on the second regime, RH$=30\%$, $T=17$°C, coarse salt, glass rod $d=16$ mm}
    \label{main_fit_20/06}
\end{figure}

For a quantitative comparison, we performed a linear fit of the experimental height data, yielding $a^\mathrm{{exp}}_2 = (0.80  \pm 0.01) \ \mathrm{cm/h}$ (see Fig.~\ref{main_fit_20/06}), compared with $a^\mathrm{{theo}}_2 = 2\pm1  \ \mathrm{cm/h}$.
The discrepancy between these values likely stems from fluctuations in the parameters used for the prediction —particularly the evaporation rate— as well as experimental noise. Nevertheless, the predicted value remains within the correct order of magnitude, confirming the overall consistency of our model.\\

\subsection{Third regime : logarithmic width}

In the third regime where the height $h$ plateaus due to gravity's effect on the flux, we expect a logarithmic time evolution for $e(z,t)$, according to Eq.~\ref{eq_epaisseur}, This behavior is confirmed experimentally in Fig.~\ref{height_width_fit20/06plusimg} (bottom).\\

Table~\ref{tab:time_caract} compares the characteristic times predicted by the model with those measured experimentally, showing good agreement in their orders of magnitude. As noted before, the discrepancies may stem from the difficulty of accurately measuring accurately the evaporation rate, as well as the overlap between regimes visible in Fig. \ref{height_width_fit20/06plusimg}.\\

\subsection{About the results of the simulation and links to observations}

\begin{figure} 
\centering
\begin{subfigure}{0.49\linewidth}
    \includegraphics[width=1\linewidth]{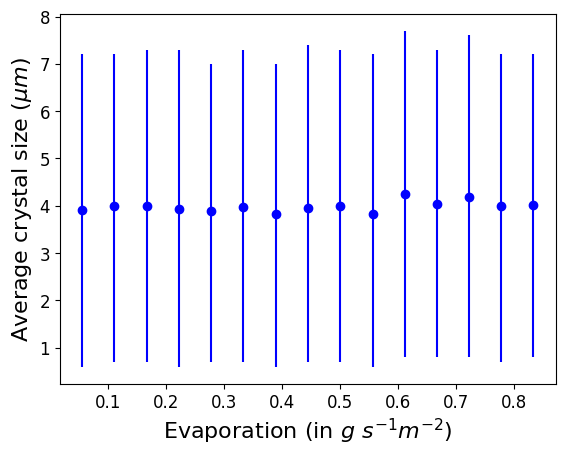}
\end{subfigure}
\begin{subfigure}{0.49\linewidth}
    \includegraphics[width=1\linewidth]{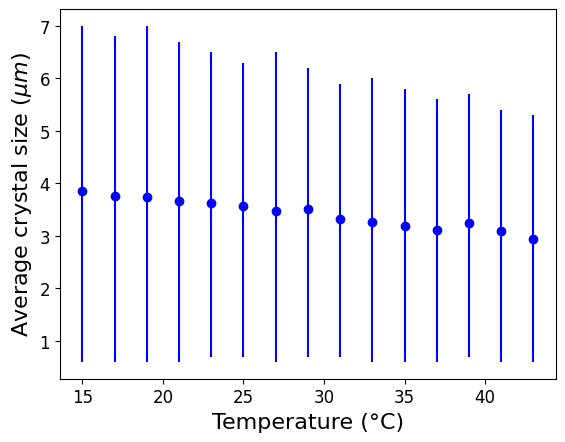}
\end{subfigure}
\caption{Evolution of the distribution of crystals sizes, with the evaporation on the left and temperature on the right, given by the simulation.}
\label{fig:simu_TJ}
\end{figure}

When the evaporation rate is varied, the simulations have a constant average and interval of confidence in Fig.~\ref{fig:simu_TJ} (left). In contrast, an increase in temperature reduces the prevalence of larger crystals, as the decrease of the 84th percentile in Fig.~\ref{fig:simu_TJ} (right) of over $1 \ \mu$m shows\footnote{In these interpretation, we suppose that the distribution is parametrized in a simple way implying that changes in the statistical measures is equivalent to changes of parameters in the model.}. This trend may explain why lowering the relative humidity affects only the crystallization kinetics but not the final size distribution, whereas increasing temperature modifies both the kinetics and the resulting crystal sizes. Moreover, higher temperatures are associated with steeper growth fronts and narrower overall width.

The simulation therefore suggests the following interpretation:
\begin{itemize}
\item A reduction in the characteristic crystal size implies smaller capillary spaces between neighboring crystals. According to Jurin’s law, this should allow water---and thus crystals---to advance to greater heights.
\item At the same time, smaller capillaries hinder the outward flow of water (i.e., increase $\xi$, as defined in Sec.~\ref{theoretical_model}), which in turn restricts the lateral spreading and keeps the width relatively small.
\end{itemize}

These observations indicate that complementary simulations and experiments at different temperatures would provide a valuable test of the macroscopic models and help refine our understanding of salt creeping dynamics.

\section{Dead Ends} 
\label{DeadEND}

In addition to our main results, we had identified promising trails for probing the phenomenon that we were not able to follow through: the mass measurement and the exploration of the structure with an optical coherence tomography apparatus. 

\subsection{Mass evolution}
 

As discussed in Sec.~\ref{discussion}, directly probing the height evolution of the first regime is relatively uneasy. 
To overcome this limitation and further test our model further, we attempted to rely on mass measurements. We monitored the weight change induced by the rod plunging in the solution.\footnote{In general, we should take into account the buoyancy force. However, the radius of the rod is negligible compared to the diameter of the container which in turn makes this force negligible.} At first order in $e(z,t)/R$ (a valid approximation in this constant-thickness-$e_0$ regime since $R \gg e_0$), and assuming an uniform average density $\rho_e$ for the wet crystal, the mass evolution is given by:
\begin{align}
    m(t) = 2 \pi R \rho_e e_0 h_0 \exp\left[\frac{\alpha \mathrm{e}^{-\xi e_0} t}{2e_0} \right].
\end{align} 

\begin{figure} 
\centering
    \includegraphics[width=0.7\linewidth]{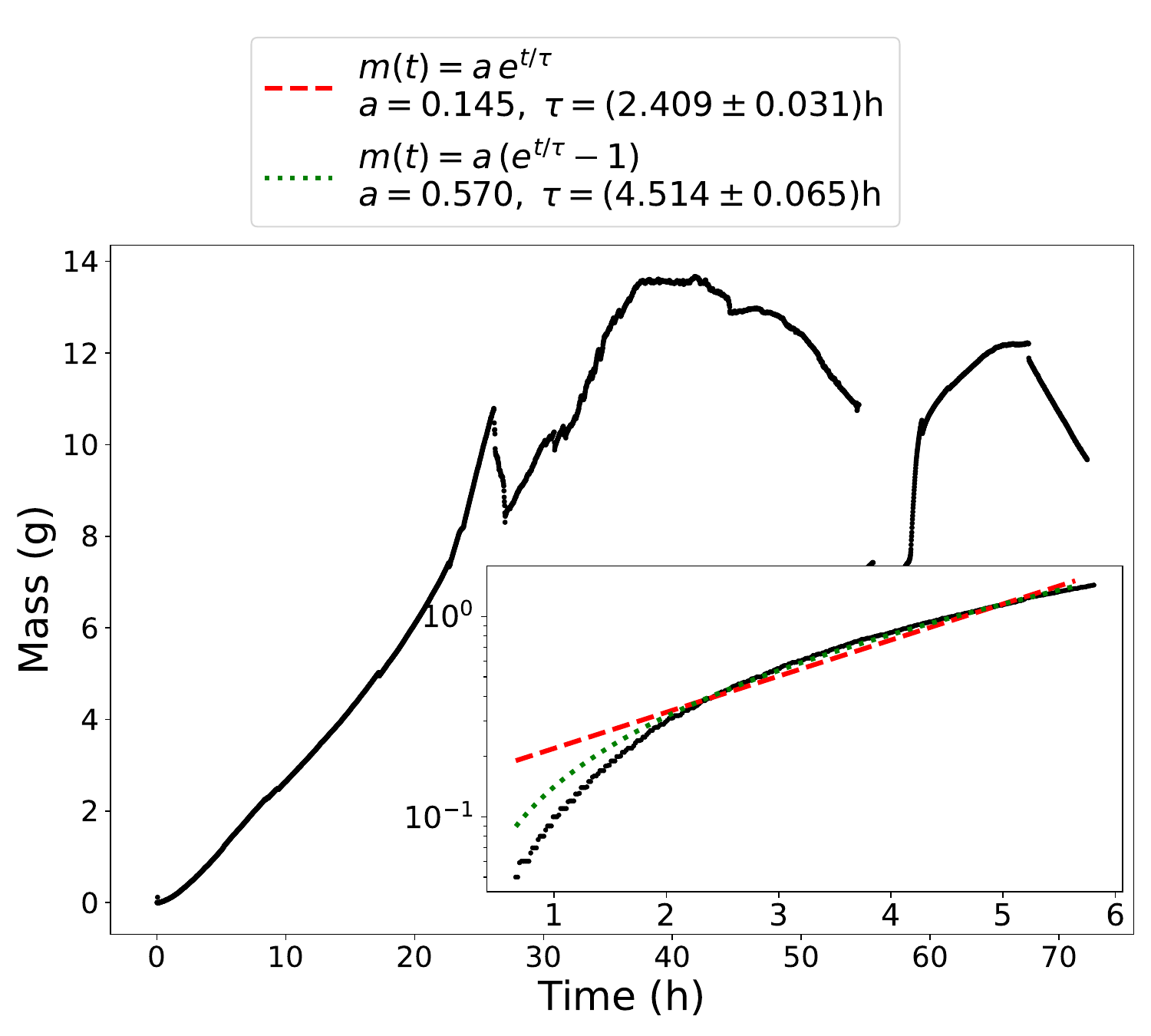}
    \caption{Mass measurement of the creeping at a relative humidity of 30\% and a temperature of 17°C.}
    \label{fit_mass_20/06}
\end{figure}

However, this prediction fits our data poorly, as shown in Fig.~\ref{fit_mass_20/06} by the red dashed line. This clash illustrates the difficulty of grasping mesoscopic phenomena using a purely macroscopic approach. Indeed, depending on the length scale considered, the initial mass is either zero (macroscopic initial condition) or driven by the wetting of the rod (mesoscopic initial condition). This fundamental difference prevents us from reaching a satisfying model for the first regime. 

It is nevertheless enlightening to observe that an exponential model, thus reflecting macroscopic initial conditions, of the form $m(t) = A[ \exp(t/\tau) -1 ]$ fits our data much better, displayed with the green dashed line in Fig.~\ref{fit_mass_20/06}. This type of model is also used in \cite{qazi2019salt}, and is typical for diffusion-limited aggregation processes. 


\subsection{Microscopic images with OCT}
The existence of a main capillary between the rod and the crystal is a central hypothesis in both our theoretical model and our simulations. First formulated by Washburn a century ago \cite{washburn}, this assumption remains debated. We tried to determine the size of said capillary along the rod using Optical Coherence Tomography -OCT is a technique based on interferences and contrast decay to recreate 3D images-. \newline

\begin{figure} 
\centering
    \includegraphics[width=0.8\linewidth]{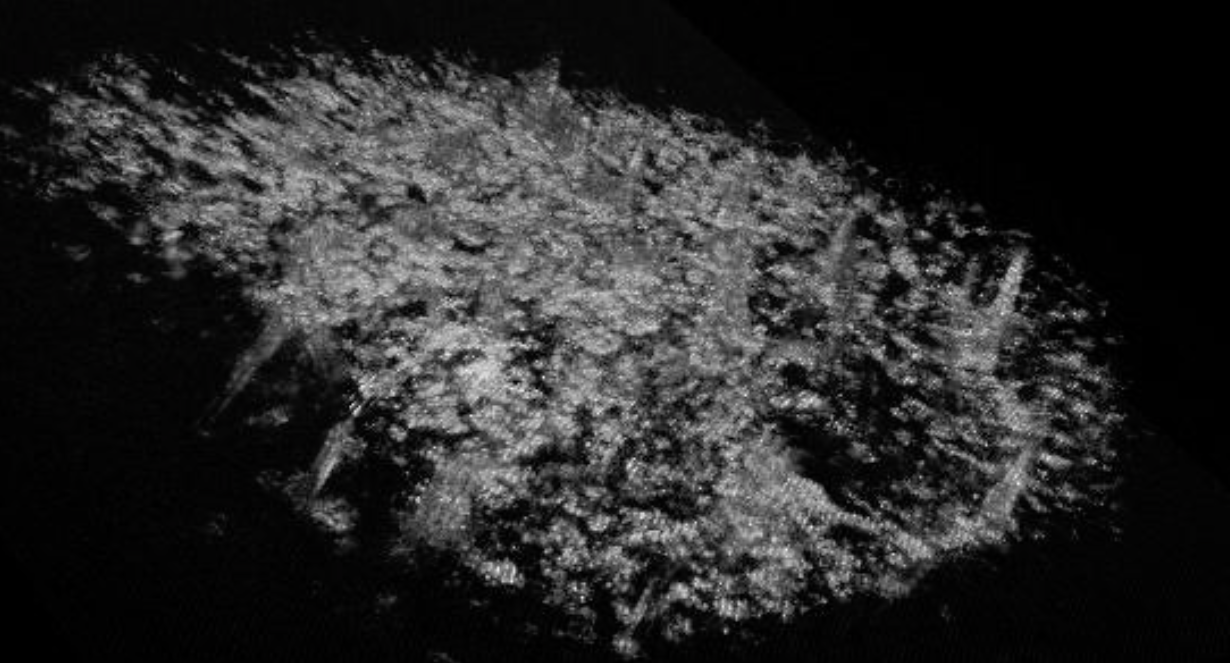}
    \caption{One point of view of the 3D image of the crystals by OCT, the diameter of the whole structure is about $370\ \mu$m. Each gray area correspond to an interface where the light can reflect.}
    \label{fig:OCT}
\end{figure}

Although we successfully observed some capillaries near the surface of the crystal -see Fig.~\ref{fig:OCT}-, we were unable to confirm the presence or measure the size of the main capillary between the rod and the crystal domain. The excessive thickness of the crystals caused significant light diffusion in the media. To mitigate such diffusion, X-ray imaging should be considered as a promising alternative.

\section{Conclusion}
Our study presents a model that successfully captures the kinetics of salt creeping on a free surface. From this model, we analytically identified three distinct regimes: an initial regime corresponding to the onset of the phenomenon, an intermediate phase where the height increases while the thickness remains constant, and a final stage where the height saturates whereas the thickness increases. In addition to being blurringly initiated, the first regime was too short for our measurement methods to precisely probe. However, the two subsequent regimes were clearly identified in the experimental data, and their numerical characteristics are consistent with the analytical predictions. While the three asymptotic regimes are accessible in our experiments, the transitions between them remain unresolved. Future investigations could address this limitation by performing continuous measurements, such as mass monitoring.

At the microscopic scale, cellular automaton simulations reproduced the main experimental trends and emphasized the significant influence of temperature. These results suggest that future investigations should focus on quantifying the temperature dependence of salt creeping (rather than relative humidity), which appears to play a key role in controlling both the microscopic and macroscopic morphology. Such studies could also provide better statistical data on the influence of these parameters ($T$ and RH) on characteristic times.

\section*{Acknowledgements}
We thank the French Physicists’ Tournament for organising the 2025 competition and for proposing this subject. We are particularly grateful to our supervisors Claire Marrache-Kikuchi, Philippe Gondret, and Maxime Varese (Université Paris-Saclay), as well as Thierry Charitat, Pierre Muller, and Maxime Dherbecourt (Université de Strasbourg), for their support during the preparation of the tournament, for providing access to their laboratories to perform experiments, and for their careful reading and valuable advice during the preparation of this manuscript. We thank also Florian Pallier of IJCLab who helped us with the microscope, as well as Marine Sebastien and Alexis Robert for their help in the Strasbourg team.

\bibliography{Supp_mat_et_biblio/biblio}

@article{qazi2019salt,
  title={Salt creeping as a self-amplifying crystallization process},
  author={Qazi, MJ and Salim, H and Doorman, CAW and Jambon-Puillet, E and Shahidzadeh, N},
  journal={Science advances},
  volume={5},
  number={12},
  pages={eaax1853},
  year={2019},
  publisher={American Association for the Advancement of Science}
}

@article{Chloride_erosion,
  title={Chloride ion erosion of pre-stressed concrete bridges in cold regions},
  author={Cui, Hongtao and Zhuo, Yi and Ke, Dongyuan and Li, Zhonglong and Li, Shunlong},
  journal={J Infrastruct Preserv Resil },
  volume={4},
  number={12},
  pages={eaax1853},
  year={2023},
  publisher={Springer Nature}
}

@book{Darcy1856,
  author    = {Darcy, Henry},
  title     = {Les Fontaines publiques de la ville de Dijon},
  publisher = {Victor Dalmont},
  address   = {Paris},
  year      = {1856}
}

@article{CEMENT1,
  title={Research and engineering application of salt erosion resistance of magnesium oxychloride cement concrete},
  author={Chang, Chenggong and An, Lingyun and Zheng, Weixin and Wen, Jing and Dong, Jinmei and Yan, Fengyun and Xiao, Xueying},
  journal={Materials},
  volume={14},
  number={24},
  pages={7880},
  year={2021},
  publisher={MDPI}
}

@article{washburn,
  title={The creeping of solutions},
  author={Washburn, Edward R},
  journal={The Journal of Physical Chemistry},
  year={1927},
  publisher={ACS Publications}
}

@article{hazlehurst,
  title={The Creeping of Saturated Salt Solutions.},
  author={Hazlehurst Jr, TH and Martin, Herbert C and Brewer, L},
  journal={The Journal of Physical Chemistry},
  volume={40},
  number={4},
  pages={439--452},
  year={2002},
  publisher={ACS Publications}
}

@phdthesis{velasco2016evaporation,
  title={{\'E}vaporation en milieu poreux en pr{\'e}sence de sel dissous. Structure et lois de croissance des efflorescences},
  author={Velasco, Mauricio Duenas},
  year={2016},
  school={Institut National Polytechnique de Toulouse-INPT}
}

@article{Ostwald-Freundlich_effect,
author = {Shchekin, Alexander and Rusanov, Anatoly},
year = {2008},
month = {11},
pages = {154116},
title = {Generalization of the Gibbs–Kelvin–Köhler and Ostwald–Freundlich equations for a liquid film on a soluble nanoparticle},
volume = {129},
journal = {The Journal of chemical physics},
doi = {10.1063/1.2996590}
}

@article{kelvin,
  author  = {Thomson, W.},
  title   = {On the Equilibrium of Vapour at a Curved Surface of Liquid},
  journal = {Philosophical Magazine},
  year    = {1871},
  series  = {4},
  volume  = {42},
  pages   = {448--452}
}

@article{jurin,
  title={II. An account of some experiments shown before the Royal Society; with an enquiry into the cause of the ascent and suspension of water in capillary tubes.},
  author={Jurin, James},
  journal={Philosophical Transactions of the Royal Society of London},
  volume={30},
  number={355},
  pages={739--747},
  year={1718},
  publisher={The Royal Society London}
}

@article{rankine_formula,
  title={The Rankine-Kirchhoff approximations for moist thermodynamics},
  author={Romps, David M},
  journal={Quarterly Journal of the Royal Meteorological Society},
  volume={147},
  number={740},
  year={2021},
  publisher={Oak Ridge National Lab.(ORNL), Oak Ridge, TN (United States). Atmospheric~…}
}

@article{concrete_attack,
author = {Zimmermann, Nils E. R. and Vorselaars, Bart and Quigley, David and Peters, Baron},
title = {Nucleation of NaCl from Aqueous Solution: Critical Sizes, Ion-Attachment Kinetics, and Rates},
journal = {Journal of the American Chemical Society},
volume = {137},
number = {41},
pages = {13352-13361},
year = {2015},
doi = {10.1021/jacs.5b08098}
}

@article{washburn2002creeping,
  title={The creeping of solutions},
  author={Washburn, Edward R},
  journal={The Journal of Physical Chemistry},
  volume={31},
  number={8},
  pages={1246--1248},
  year={2002},
  publisher={ACS Publications}
}

@article{application-creeping-solar,
  title={Self-assembled porous salt crystals for solar-powered crystallization},
  author={Yu, Jie and Zhang, Lenan and Gao, Jintong and Han, Wenyu and Wang, Ruzhu and Xu, Zhenyuan},
  journal={Energy \& Environmental Science},
  volume={18},
  number={1},
  pages={454--467},
  year={2025},
  publisher={Royal Society of Chemistry}
}

@article{application-creeping-solar-2,
  title={A comprehensive review of salt rejection and mitigation strategies in solar interfacial evaporation systems},
  author = {Liu, Shang and Yang, Qijun and Li, Shiteng and Lin, Meng},
  journal={Desalination Vol. 600
p. 118507
},
  volume={600},
  pages={118507},
  year={2025}
}

@book{kittel2005,
  author    = {Charles Kittel},
  title     = {Introduction to Solid State Physics},
  edition   = {8},
  publisher = {John Wiley \& Sons},
  address   = {Hoboken, NJ},
  year      = {2005},
  isbn      = {978-0471415268}
}

\end{document}


\title{Supplementary material to : Kinetics of Salt Creeping on a Free Surface: From Nucleation to Saturation }

\author{\firstname{Baptiste} \lastname{Guilleminot}\inst{1}\fnsep\thanks{\email{baptiste.guilleminot@gmail.com}} \and
        \firstname{Élodie} \lastname{Harlé}\inst{2}\fnsep\thanks{\email{elodie.harle67@gmail.com}} \and
        \firstname{Timothée} \lastname{Herbeau}\inst{1}\fnsep\thanks{\email{herbeautim@yahoo.com}}
}

\institute{Université Paris-Saclay, FRANCE
\and
           Université de Strasbourg, 67000 Strasbourg, FRANCE
}

\date{\today}

\abstract{The phenomenon of salt creeping on a free surface remains only partially understood, particularly with respect to its dynamics. In this work, combining a theoretical model with controlled experiments, we identify three distinct kinetic regimes: an initial exponential growth followed by a subsequent linear regime, and a final stage where the height saturates while the crystal deposit thickens logarithmically. This unified description makes it possible to follow the macroscopic kinetics of salt growth on a free surface from its nucleation to saturation. In addition, we complement this macroscopic analysis with numerical simulations that shed light on the evolution of the microscopic crystal structure under varying external conditions (humidity and temperature).
}

\maketitle

\section{Model for the flux}

\label{model_flux}

\begin{figure} 
\centering
    \includegraphics[width=1\linewidth]{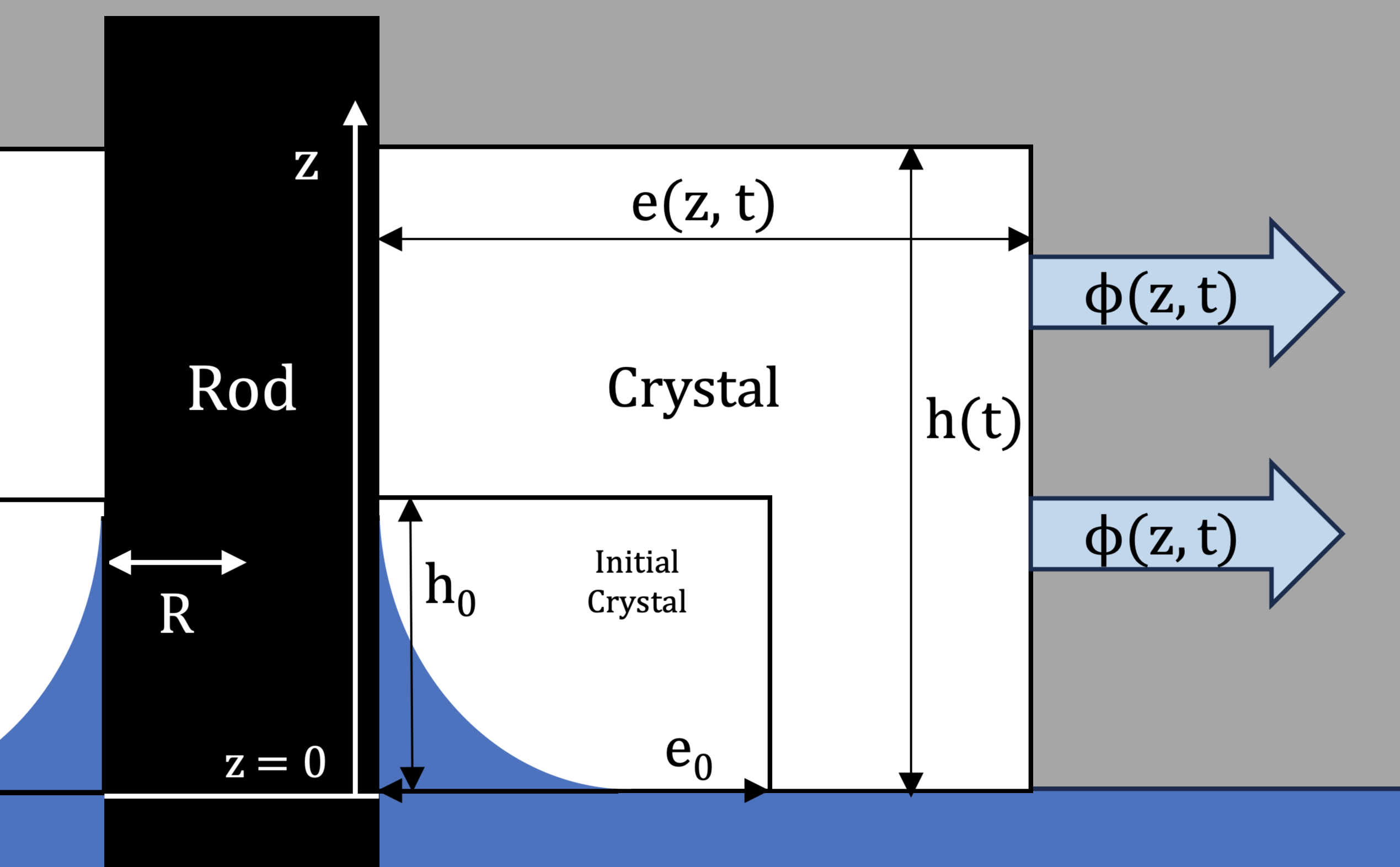}
    \caption{Definition of the different variables used, the evaporation flux $\phi(z,t)$ of water molecules from the crystal layer, of thickness $e(z,t)$ and height $h(t)$. The first crystal nucleation occurs at the meniscus of brine on the rod of radius $R$. Hence, the meniscus parameters $h_0$ and $e_0$ determine the initial crystal domain.}
    \label{schema_variable}
\end{figure}

Based on experimental observations (Fig.\ref{schema_variable}), the crystalline layer forming around the rod is modeled as a porous medium composed of an assembly of salt crystals with approximately uniform micrometric size. These crystals are separated by interstitial spaces —capillaries— through which the saturated salt solution flows radially before evaporating at the outer surface. At a given altitude along the rod, the solution is supplied predominantly through the main capillary located between the rod surface and the crystalline layer. Due to the rotational symmetry of the system, cylindrical coordinates are used throughout the paper.\newline 

To derive an expression for the flux of salted water $\Phi(e,z,t)$ evaporating at the surface of the crystalline layer, we first model the radial flux of liquid salted water 
$\phi(r,z,t)$ rising by capillarity from the solution through the crystal. The latter describes the transport of water from the reservoir to a height $z(t)$ within the porous crystal.\newline

At the outer edge of the crystalline layer, the flux of water vapor leaving the crystal equals the flux of liquid water transported through it:
$\Phi(e,z,t) = \phi(r=R+e,z,t) $.\\

As noted in the main text, this model is somewhat unconventional. In typical porous media, fluid flow is usually described using Darcy’s law. However, this approach is insufficient here, because it does not account for the crystallization of $\mathrm{NaCl}$  as the solution moves through the crystal layer.\newline

By discretizing the crystal thickness $e(z,t)$ into $N$ slabs of equal height, we account for the progressive decrease of the water flux as it passes through the layer. This reduction occurs because $\mathrm{NaCl}$ molecules are deposited along the path, contributing to the growth of existing crystals. Assuming each slab contributes equally, with $\xi$ denoting the number of $\mathrm{NaCl}$ molecules deposited per unit length, we obtain the following equation:

\begin{align}
       \frac{\phi_{n+1}}{\phi_n} = 1\ - \xi \frac{e(z,t)}{N}.   
\end{align} 
One can then express $\phi_n$ as a function of $e(z,t)$ and a constant $ C$ which we interpret physically below.
\begin{align}
       \phi_n = C\ \left(1 - \frac{\xi e(z,t)}{N} \right)^n.\
\end{align} 

\noindent Finally, taking the limit $N \to \infty$ of $\phi_{n=N}$ gives the water flux in the continuous limit:
\begin{align}
    \phi_{n=N} &= \lim_{N\to \infty}C\ \left(1 - \frac{\xi e(z,t)}{N} \right)^N \nonumber\\ &=  \lim_{N\to \infty}C\ \exp\left(N \ln\left[1 - \frac{\xi e(z,t)}{N}\right] \right) \nonumber \\ 
      &= C \exp\left( -\xi  e(z,t)\right).
\end{align}

This should equate the flux of water evaporating at the surface of the salt layer:
 
\begin{align}\label{Flux_1}
        &\phi(r=R+e(z,t)) = \lim_{N\to \infty} \phi_{n=N},\nonumber \\
        &\Phi(e,z,t) = \phi(r=R+e(z,t)) = C\ e^{-\xi  e(z,t)}.   
\end{align} 

It is now clear that the multiplicative constant should be the surface evaporation of brine at given temperature and pressure conditions: $C = \Phi_0$.
Moreover, we  interpret the quantity $\xi $ as the inverse of $e_{\infty}$, the characteristic length above which flux is significantly reduced.

The evaporation flux decreases as the crystalline layer thickens, and the capillaries shrink. However, gravity also affects the flux and we must modify Eq.\ref{Flux_1} to account for this.\newline
Noting the very slow evolution of the system compared to fluid mechanics timescales, we use equilibrium statistical physics and the usual Boltzmann weight to represent the fraction of the total $\mathrm{NaCl}$ concentration available at a given altitude $z$. Explicitly, the effective concentration of NaCl molecules in the liquid present at an altitude $z$ is $C_0e^{-z/h_\infty}$ where $C_0$ is the saturated concentration in salt, found at $z=0$. We argue that this effective concentration leads to an effective evaporation flux $\Phi_0 \ e^{-\frac{z}{h_\infty}}$. The quantity $h_\infty$ should be interpreted as the characteristic length above which gravity hinders the flux by notably decreasing the concentration in $\mathrm{NaCl}$. Finally, our model reads:
 
\begin{align}\label{Flux_supmat}
    \Phi(e,z,t) = \Phi_0 \ e^{-\frac{z}{h_\infty}}  \ e^{-\xi \ e(z,t)}.
\end{align}
 
In the following we retain the experimental value $h_\infty \approx 0.1 \ \mathrm{m}$ in order to circumvent the substantial uncertainties associated with the wetting angle, the capillary size, and the value of the surface tension of salt-saturated water. It is important to note that, although the experimental value of $h_\infty$ is used as an input parameter of the theoretical model predictions, it is expected to be independent of the kinetics of salt creeping. This height is determined solely by physical constants and hydrostatic equilibrium.

\section{Derivation of the three regimes}
\label{theoretical_model}

We have derived an analytical model for the time evolution of the height $h(t)$ and width $e(z,t)$ of the crystal domain on the rod of radius R (2D cylindrical invariance is assumed). These variables and the growth process are illustrated in Fig.~\ref{schema_variable}.

Our model is based on an equation expressing the particle conservation -both water molecules and $\mathrm{NaCl}$- throughout evaporation and crystallization. Expanding in the approximation $e(z,t) \ll R$ the volume $(R + e(z,t))^2 - R^2 \approx 2R\ e(z,t)$.

\begin{align}
       &\frac{d}{dt} \int_{0}^{h(t)}dz \int_{0}^{2 \pi} d\theta\int_{R}^{R+e(z,t)} rdr \\ \nonumber &= \frac{\langle V \rangle}{N_e N_s} \int_{0}^{h(t)}dz \int_{0}^{2 \pi} \Phi(e,z,t) (R+e(z,t)) d\theta.
\end{align}

\noindent Which we can further develop:
\begin{align} 
       &\frac{d}{dt} \int_{0}^{h(t)}dz \ 2 \pi \frac{(R+e(z,t))^2-R^2}{2} \\ \nonumber &= \frac{\langle V \rangle}{N_e N_s} \int_{0}^{h(t)}dz \ 2 \pi (R+e(z,t)) \ \Phi(e,z,t) .
\end{align}

\noindent Finally, in the limit $e(z,t) \ll R$: 

\begin{align}\label{Main_equation}
        &\frac{d}{dt} \int_{0}^{h(t)}dz \  e(z,t) \\ \nonumber &= \frac{\langle V \rangle}{N_e N_s} \int_{0}^{h(t)}dz \   \Phi(e,z,t) .
\end{align}

\noindent Where $\langle V \rangle$ denotes the average volume of a single crystal unit and $N_s$ the number of entity $\mathrm{NaCl}$ in a volume $\langle V \rangle$.

Moreover, we let $N_e$ the ratio of the number of NaCl entities to water molecules in the solution. It is fixed by the temperature and pressure, given that the solution is saturated in salt. 

Finally, $\Phi $ is the flux, per unit surface, of water molecules evaporating at the surface of the crystal (in $\mathrm{s}^{-1} \mathrm{m}^{-2}$) for a given altitude and time.
In the section \ref{model_flux}, we showed:
\begin{align}
     \Phi(z,t) = \Phi_0 \ e^{-\frac{z}{h_\infty}}  \ e^{-\xi \ e(z,t)}.  
\end{align}
At a given temperature and pressure, the evaporation rate of water is denoted $\Phi_0$. 

With Reynold's transport theorem, this equation can be rewritten as the following:

\begin{align}\label{kinetic}
      &\frac{d}{dt} h(t) \  e(h(t),t)  \\ \nonumber &+ \int_{0}^{h(t)}  \partial_t e(z,t) \ \ dz =  \frac{\langle V \rangle}{N_e N_s} \int_{0}^{h(t)} \Phi(e,z,t) \ dz. 
\end{align}

The defining limits of the first regime are:
 
\begin{align}
    \partial_t e(z,t)dz = \dot h \delta(z&=h(t)) e_0, & 
    \frac{h(t)}{h_\infty}  \ll 1.
\end{align}

Denoting  $\alpha = \Phi_0 \langle V \rangle/(N_e N_s)$ and injecting this observed behaviour in our conservation equation leads to:

\begin{align}\label{1and2regime}
        \dot h &= \frac{\alpha}{2e_0 \Phi_0} \int_{0}^{h(t)} \Phi(e,z,t) \ dz    = \frac{\alpha}{2e_0}h_{\infty}\ e^{-\xi e_0} \left(  1 - e^{-\frac{h(t)}{h_\infty}}     \right),
\end{align} 
  
\noindent With $\alpha = \Phi_0 \langle V \rangle/(N_e N_s)$ (in $\mathrm{m}. \mathrm{s^{-1}}$). Hence, in the limit of the first regime $h(t)/h_\infty \ll 1 $ we develop the second term and solve: 

\begin{align}
       h(t) = h_0 \ \exp[ \ {\frac{\alpha e^{-\xi e_0}}{2e_0} t} \  ].
       \label{eq_hauteur}
\end{align}

\noindent With $h_0$ the initial height of the crystal formation. Since the phenomenon is started by the wetting of the rod, $h_0$ will be identified to this height. Similarly, $e_0$ is the thickness of the meniscus.

Looking at the converse set of hypothesis: 
\begin{align}
    \partial_t e(z,t)dz  = \dot h \delta(z&=h(t)) e_0, & 
    \frac{h(t)}{h_\infty} = \mathrm{O}(1).
\end{align}
 
\noindent In Eq.\ref{1and2regime} (which was derived only assuming constant crystal thickness if any), these limits lead to the following equation: 

\begin{align}
        \dot h &= \frac{\alpha}{2e_0}h_{\infty}\ e^{-\xi e_0 }.  
\end{align} 

\noindent This in turns gives a linear evolution for $h(t)$:

\begin{align}
    h(t) = h_0 + \frac{\alpha}{2e_0}h_{\infty}\ e^{-\xi e_0} \ t. 
    \label{eq_linear}
\end{align}
With $h_{\infty} = 0.1\ \mathrm{m}$ as discussed above.

Finally, the third regime, where $h(t)/h_{\infty}  \sim1$ saturates, leads to a prediction for the thickness evolution $e[z, t]$. The first term in the left hand side of Eq.\ref{kinetic} vanishes. A sufficient condition on the field $e(z,t)$ for the equation to be satisfied is :

\begin{align}
      \frac{\langle V \rangle}{2N_e N_s}\Phi(e(z,t))  - \partial_t e(z,t) = 0. 
\end{align} 

\noindent Combined with our model for the flux -Eq.\ref{Flux_supmat}- gives the ODE: 

\begin{align}
      \alpha e^{-\xi e(z,t)} e^{-\frac{z}{h_\infty}}  - \partial_t e(z,t)   =0.
\end{align} 

\noindent We can solve the latter, at fixed $z$, up to a constant $C(z)$:
\begin{align}
        e(z,t)  = \frac{1}{\xi} \ ln\left(C(z) + \xi \alpha e^{-\frac{z}{h_\infty}} \ t \right)    
        \label{eq_epaisseur}.
\end{align}
\noindent Which corresponds to the logarithmic evolution provided in the main text.

\section{Modelling of capillary rise above the uppermost crystal}

For salt to creep upward, water must rise slightly above the highest crystal. This occurs through the formation of a small meniscus between the uppermost crystal and the rod. A fully consistent description would require accounting for the wetting angles of water on both salt and glass, which are constrained by the surface tension balance at the three-phase contact line.

However, introducing this level of detail would add complexity disproportionate to the rest of the model. Since the uppermost crystals are small, the meniscus geometry is well approximated by a linear relation between its vertical extent $a$ and the characteristic crystal size 
$b$. Based on microscopic observations, we set $a = 3b$.

\section{Construction of the crystallization probability distribution}

The probability of crystallization is determined by the local salt concentration and the energetic cost of forming a new crystal cell. In this section, we derive the concentration field from the evaporation dynamics and the Ostwald--Freundlich relation before combining it with a Boltzmann factor to obtain the probability distribution used in the simulations. 

The concentration of NaCl at the crystal-water interface of a crystal with characteristic size $\tilde r$ is taken from the Ostwald-Freundlich relation :

\begin{equation}
    \frac{C^{\tilde r}_{\rm sat}}{C^\infty_{\rm sat}}
    = \exp\left(\frac{E_{\rm ts}(\tilde r)}{RT}\right)
    \label{eq:ostwald}
\end{equation}

where $E_{\rm ts}(r) = \frac{2\gamma V_{\rm m}}{r}$ is the surface tension energy, $R$ is the ideal gas constant, $\gamma$ is the surface tension between the glass and the brine and $V_{\rm m}$ is the volume per mole of NaCl. Here, we assume that the crystal is spherical with a characteristic size $r$, which constitutes a first-order approximation. From the crystal surface the concentration is assumed to vary with distance $z$ (measured tangent to the rod) according to a one-dimensional steady balance explained in the next section:

\begin{equation}
    \frac{C}{C_0} = \exp\left(\sqrt{\frac{J}{\rho r D}}z\right)
\end{equation}

where $\rho$ is the density of water and $D$ is the diffusion constant of NaCl in water. With these two equations, we may know the concentration at a distance $z$ of a crystal of size $\tilde r$.

Now, we need a model for the probability of crystallisation. We assume that crystallization events occur close to local quasi-equilibrium, so that the probability of a crystallization event in a cell during one time step follows a Boltzmann factor. Let $\Delta E$ be the change in free energy associated with converting the liquid cell (volume $V_{\mathrm{cell}}$) into a crystalline cell. We write

\begin{equation}
    P \propto \exp \left(-\beta\Delta E\right).
\end{equation}

In our discrete, lattice-based approximation, $\Delta E$ is decomposed into an interfacial (bond) contribution arising from creating new crystal--crystal links with neighboring crystalline cells and a bulk term. For a candidate cell sharing $n$ sides (neighbors) with existing crystal we take

\begin{equation}
    \Delta E = -\frac{n l E_{\rm l}}{3dx} + E_{\rm vol}
\end{equation}

where $l$ is the bond length, $E_{\rm l}$ the energy per bond, $\Delta x$ the grid spacing, and $E_{\mathrm{vol}}$ the bulk crystallization energy for the cell. Because $E_{\mathrm{vol}}$ is common to all candidate cells it cancels during normalization of the probabilities. The direct contribution of the surface-tension term is neglected because, in our parameter range, it is three orders of magnitude smaller than the bond term $n l E_{\rm l} /(3 \Delta x)$. At very small $\Delta x$, the Ostwald--Freundlich effect diverges the model no longer applies.

The nucleation time is of the order of seconds to minutes. The evaporation, the concentration and the length of the phenomenon gives us a characteristic time of a couple of hours. Therefore we should use the latter to gives us the volume of crystal between time steps. We use that information to normalize the probability. With all of that, a random draw is done to determine where the new crystals will be placed. 

\section{Evolution of the Concentration Field}

We assume that the system is in a quasi-steady regime. This approximation is justified because the characteristic diffusion time of salt in water (of order a few seconds over micrometric distances) is much shorter than the characteristic timescale of the creeping process (hours). The concentration field can therefore be treated as stationary.

The transport of dissolved NaCl is governed by the steady advection--diffusion equation
\begin{equation}
\nabla \cdot (C \vec{v}) = D \nabla^2 C,
\end{equation}
where $C$ is the NaCl concentration, $D$ its diffusion coefficient in water, and $\vec{v}$ the velocity field of the liquid phase.

We assume that transport occurs predominantly along a single vertical direction $z$, measured upward from the uppermost crystal ($z = 0$). The liquid film is assumed to have a constant thickness $h$, comparable to the size of the last crystal. Under these assumptions, the equation reduces to
\begin{equation}
\partial_z (v_z C) = D \, \partial_z^2 C.
\end{equation}

The vertical velocity gradient is constrained by evaporation and incompressibility. If $J$ denotes the mass evaporation flux (kg$\cdot$m$^{-2}\cdot$s$^{-1}$) and $\rho$ the density of water, mass conservation gives
\begin{equation}
\partial_z v_z = \frac{J}{\rho h}.
\end{equation}

Substituting this relation into the advection--diffusion equation yields
\begin{equation}
\partial_z^2 C
=
\frac{J}{\rho h D} \, z \, \partial_z C
+
\frac{J}{\rho h D} \, C.
\end{equation}

We introduce the characteristic length
\begin{equation}
z_0 = \sqrt{\frac{\rho h D}{J}} \approx 180 \,\mu\mathrm{m}.
\end{equation}

Since the relevant distances in our system are of order a few to a few tens of micrometers ($z \ll z_0$), the term proportional to $z \, \partial_z C$ can be neglected. The equation then simplifies to
\begin{equation}
\partial_z^2 C = \frac{1}{z_0^2} C,
\end{equation}
whose solution is an exponential profile. Imposing $C = C_{\mathrm{sat}}$ at $z=0$ gives
\begin{equation}
\frac{C}{C_{\mathrm{sat}}}
=
\exp\left(\frac{z}{z_0}\right).
\end{equation}

This expression is the concentration profile used in the main text.